\documentclass[pre,aps,preprint,epsf]{revtex4}
\usepackage{amsmath,epsfig,slashbox}
\input epsf

\begin{document}
\bibliographystyle{apsrev}

\title{Fluid Permeabilities of Triply Periodic Minimal Surfaces}

\author{Y. Jung$^{1,2}$ and S. Torquato$^{1,2,3}$}
\affiliation{
$^1$Princeton Institute for the Science and Technology of Materials, Princeton 
University, Princeton, New Jersey 08544 \\
$^2$Department of Chemistry, Princeton University, Princeton, New Jersey 08544\\
$^3$Program in Applied and Computational Mathematics, Princeton University, 
Princeton, New Jersey 08544}

\date{\today}

\begin{abstract}
It has recently been shown that triply periodic two-phase bicontinuous composites
with interfaces that are the Schwartz primitive (P) and diamond (D)  minimal surfaces are not only 
geometrically extremal but extremal for simultaneous transport of heat and electricity.
The multifunctionality of such two-phase systems has been further established by demonstrating 
that they are also extremal when a competition is set up between the
effective bulk modulus and electrical (or thermal) conductivity of the 
bicontinuous composite. Here we compute the fluid permeabilities of these and other 
triply periodic bicontinuous structures  at a porosity $\phi=1/2$ using 
the immersed boundary finite volume method. The other triply periodic porous media that we 
study include the Schoen gyroid (G) minimal surface, two different pore-channel models, and an 
array of spherical obstacles arranged on the sites of a simple cubic lattice. We find that 
the Schwartz P porous medium has the largest fluid permeability among all of the six triply periodic porous
media considered in this paper. The fluid permeabilities are shown
to be inversely proportional to the corresponding specific surfaces for these
structures. This leads to the conjecture that the maximal fluid permeability for
a triply periodic porous medium with a simply connected pore space
at a porosity $\phi=1/2$ is achieved
by the structure that globally minimizes the specific surface.

\end{abstract}
\maketitle

\section{Introduction}
\label{section1}

\begin{figure}[h!]
\vspace{0.1in}
\centerline{
\epsfxsize=2.1in \epsffile{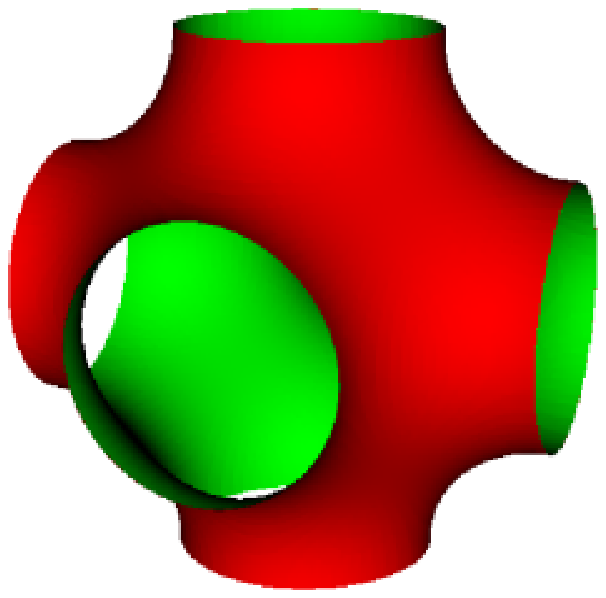} \hspace{0.0in}
\epsfxsize=2.1in \epsffile{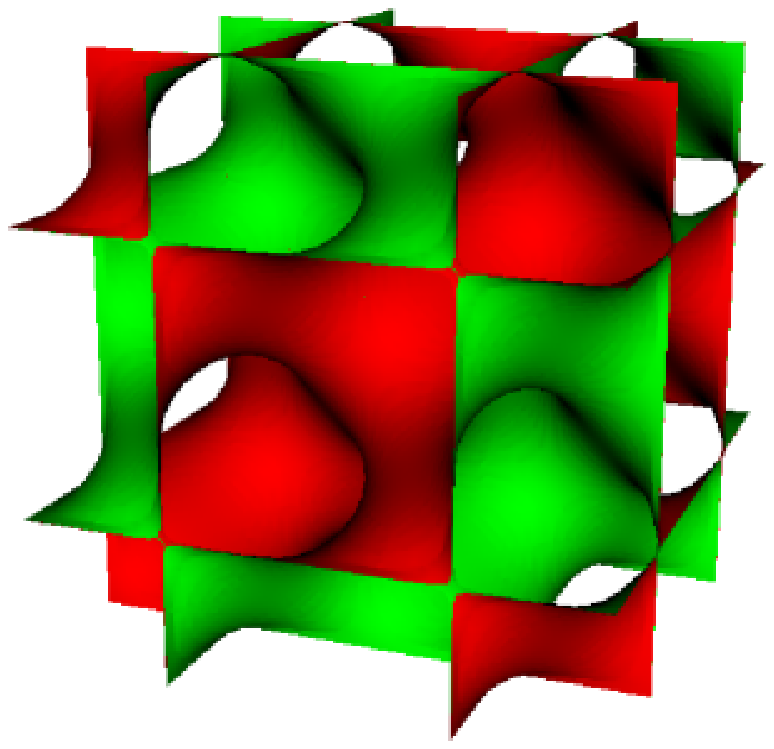} \hspace{0.0in}
\epsfxsize=2.1in \epsffile{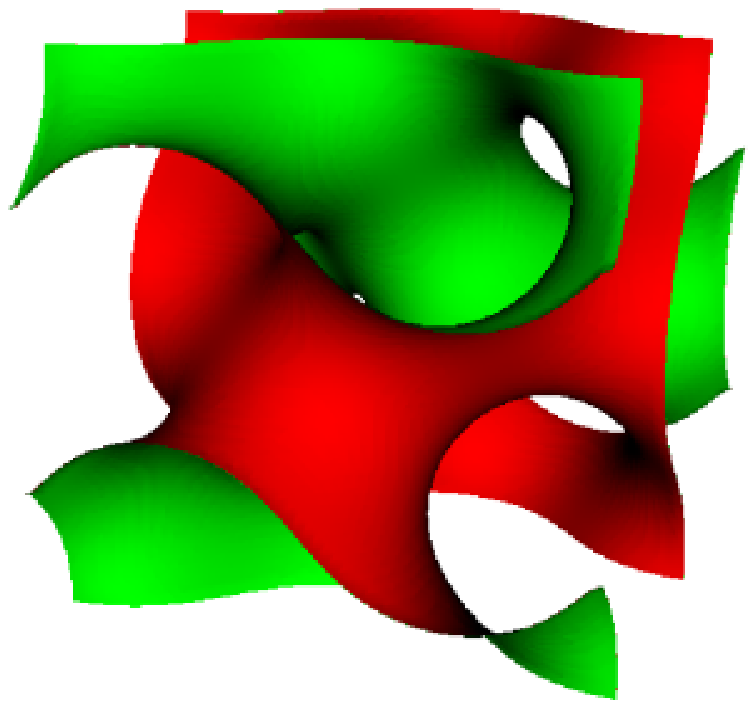} 
}
\caption{(Color online) Unit cells of three different minimal surfaces.
Left panel: Schwartz primitive (P) surface. 
Center panel: Schwartz diamond  (D) surface.
Right panel: Schoen gyroid (G) surface.}
\label{fig:tpms}
\end{figure}

Triply periodic minimal surfaces are objects of great interest to physical
scientists, biologists, and mathematicians. A minimal surface, such as a soap film, 
is one that is locally area-minimizing. Minimal surfaces are defined as surfaces with zero mean curvature.
A remarkable class of  minimal surfaces are those 
that are triply periodic (i.e., periodic in three directions).
A triply periodic minimal surface is infinitely extending, has one of the crystallographic space groups as
its symmetry group and, if it has no self-intersections, it partitions space into two disjoint but
intertwining regions  that are simultaneously continuous \cite{An90}.
If one thinks of each of these disjoint regions
as the phase of a composite, then those two-phase composites
in which the interface is a triply periodic minimal surface are
a special class of bicontinuous two-phase composites. 
The Schwartz primitive (P), the Schwartz diamond (D) and the Schoen gyroid (G) minimal
surfaces partition space into two disjoint but congruent regions, i.e., the
volume fraction of each phase is $1/2$ (see Fig.~\ref{fig:tpms}).
Triply periodic minimal surfaces arise in a variety of systems, including block 
copolymers \cite{Ol98}, nanocomposites \cite{Br01}, micellar materials 
\cite{Zi00}, and lipid-water systems and certain cell membranes 
\cite{Gel94,Land95,Klin96,Na96}.

One of us has very recently shown that triply periodic two-phase bicontinuous composites
with interfaces that are the Schwartz P and D  minimal surfaces 
are not only geometrically extremal but extremal for simultaneous transport of heat and
electricity \cite{To02b,To03}. More precisely, these are
the optimal structures when a weighted sum of the effective thermal and electrical
conductivities is maximized for the case in which
phase 1 is a good thermal conductor but poor electrical
conductor and  phase 2 is a poor thermal conductor but good electrical
conductor. The demand that this sum is maximized sets up a {\it competition}
between the two effective transport properties, and this demand
is met by the Schwartz P and D structures.
We note that for reasons of mathematical analogy, the optimality
of these   bicontinuous composites applies to any
of the pair of the following scalar effective properties:
electrical conductivity, thermal conductivity,
dielectric constant, magnetic permeability, and diffusion
coefficient. 

Since triply periodic minimal surfaces
arise in certain cell membranes, it stands to reason 
that such structures must also be relatively stiff
elastically in order to withstand mechanical loads. The topological property
of bicontinuity of these structures suggests that they would
be mechanically stiff even if one of the phases is a
compliant solid or a liquid, provided that the other phase
is a relatively stiff material. Motivated by this biological
example, Torquato and Donev \cite{To04} asked what are the two-phase
composite structures that maximize a weighted sum of the dimensionless effective
bulk modulus and effective electrical (or thermal) conductivity. Thus, they set up
a competition between a mechanical property and a transport property.
They demonstrated that triply periodic two-phase bicontinuous composites with interfaces
that are the Schwartz P and D minimal surfaces are again optimal, further establishing 
the multifunctionality of such two-phase systems.

Based on the aforementioned considerations, it is reasonable 
to inquire whether the minimal surfaces are optimal for
other effective properties, such as flow characteristics.
The purpose of this paper is to study Stokes (slow viscous) flow through triply porous
media whose interfaces are the Schwartz P, the Schwartz D, and the Schoen G
minimal surfaces. In particular, we are interested in computing the
fluid permeability $k$, which is the key macroscopic property for describing 
Stokes flow through porous media \cite{Jo86,Fe87,To02a,Sa03}. The quantity $k$ 
is the
proportionality constant between the average fluid velocity and applied
pressure gradient in the porous medium as defined by Darcy's law:
\begin{equation}\label{eq:darcy}
\bar{U}\ =\ -\frac{k}{\mu} \nabla p\ ,
\end{equation}
where $\bar{U}$ is the average fluid velocity, $\nabla p$ is the applied
pressure gradient and $\mu$ is the dynamic viscosity. 
Our expectation is that the Schwartz P porous medium will
have the highest fluid permeability among the three minimal
surfaces because it has the smallest {\it specific surface} (i.e., interface area per unit volume).
Surface area alone of course does not necessarily inversely correlate
with fluid permeability \cite{Jo86,To02a}. We will show that the 
Schwartz P porous medium does indeed have the highest fluid permeability among the three minimal
surfaces. A natural question to ask is whether the fluid permeability
of the Schwartz P porous medium is maximal in any sense. Although
we do not answer this challenging question here, we also compute  fluid permeabilities 
for other triply periodic geometries,  such as flow over
spheres arranged on the sites of a simple lattice and  flow inside two different
pore-channel models at a porosity (volume fraction of
pore space) $\phi = 1/2$.
We find that the Schwartz P porous medium has the highest
fluid permeability among all of the six triply periodic porous
media considered in this paper. We demonstrate that the fluid permeabilities are 
inversely proportional to the corresponding specific surfaces for these
structures. This leads to the reasonable conjecture that the maximal fluid permeability
(scaled by the cell length squared) for
a triply periodic porous medium with a simply connected pore space
at a porosity $\phi=1/2$ is achieved
by the structure that globally minimizes the specific surface.

Due to the numerical difficulties in generating complex meshes of various
minimal surfaces, the immersed body finite volume method was adopted in
this paper \cite{Jkim01,Fadl00}. Momentum forcing is applied on the body
surface or inside the body to satisfy the no-slip boundary condition on the
immersed boundary and also to satisfy the continuity for the each control volume
element containing the immersed body. This procedure is based on
a finite-volume approach on a staggered mesh with a fractional step method.
The unit domain is chosen to be a cube and is digitized into small cubic 
subvolumes.

In Section \ref{section2}, we review some procedures to represent 
triply periodic minimal surfaces.
In Section \ref{section3}, the immersed-boundary finite volume method 
for flow through triply periodic porous media is  described. 
In Section \ref{section4}, we verify the accuracy of our numerical
methods by computing the fluid permeability for flow over
spheres arranged on the sites of a simple cubic and compare them
to corresponding  analytical/numerical results of Sangani and Acrivos \cite{Sang82}. 
Here we also present our evaluations of the  fluid permeabilities 
for other triply periodic porous media, including Schwartz P and  D surfaces and 
the Schoen G surface and two different pore-channel models.
Concluding remarks and a discussion are given in Section \ref{section5}. 

\section{Triply Periodic Minimal Surfaces}
\label{section2}

Triply periodic minimal surfaces can be characterized exactly using an 
Enneper-Weierstrass 
(complex integration) representation only for several cases. 
For Schwartz P and D and Schoen G, analytic surfaces are constructed in 
terms of elliptic functions by the Enneper-Weierstrass 
representation \cite{Gand99,Gand00a,Gand00b}, but in practice this is 
difficult to use numerically. Recently, it was shown that
triply periodic minimal surfaces can be generated as the local minima of the scalar order parameter 
Landau-Ginzburg functional for describing ordering phenomena in microemulsion 
\cite{Gozd96a}. This Landau-Ginzburg free-energy functional 
\cite{Gozd96a} can be utilized to calculate numerically a discretization of a 
potential $\psi(x)$ such that $\psi(x)>0$ for points in phase 1 and 
$\psi(x)<0$ for points in phase 2. At the phase interface, $\psi(x)=0$, which 
in this case is a Schwartz P,D or Schoen G minimal surface. 
Moreover, the minimal surfaces can be approximated by the periodic nodal surface 
(PNS) of a sum defined in terms of the Fourier series \cite{Schn91,Mack93,Schw99,Gand01},
because any periodic surface can be expressed as the sum of an infinite number
Fourier terms.

In this paper, the Schwartz P and D and the Schoen G surfaces are produced
by both the Landau-Ginzburg minimization \cite{Gozd96a} and PNS approximations 
with a few Fourier terms \cite{Schw99,Gand01} on a $129^3$ unit cell. 
>From these potentials, one can readily make
discretizations of triply periodic bicontinuous two-phase composites in
which the two-phase interface is a minimal
surface.  Because phase 1 is topologically and geometrically identical
to phase 2, either phase can
be chosen to be the fluid phase. Indeed, triply periodic minimal surfaces
have the special property that the fluid permeability of phase 1 is
equal to the fluid permeability of phase 2. Such phase interchange
relations are rare in the case of the fluid permeability. By contrast,
phase interchange relations in the case of the effective conductivity
and effective elastic moduli of two-phase composites have been known
for a long time; see \cite{Gi96,To02a,Mi02} and references therein.
We then used an immersed boundary
finite-volume method (described below) to calculate numerically the fluid permeabilities of 
these composites.

\section{Numerical Simulation for Stokes Flow through Triply Periodic Porous 
Media}
\label{section3}

We consider the flow field in an infinite porous medium. The Navier-Stokes 
equations are nondimensionalized with $t^\prime := \mu t / (\rho L^2)$ and 
$p^\prime := (p - p_0)/(\mu U/L)$ where $t$ is the time, $\mu$ is the dynamic 
viscosity, $\rho$ is the density, $L$ is the characteristic length, $p$ is the 
pressure, $p_0$ is the pressure at a reference, $U$ is the characteristic 
speed. Dropping the primes leads to the dimensionless Navier-Stokes equations
\begin{align}\label{eq:navier-stokes}
\begin{split}
\frac{\partial u_i}{\partial t}+ Re \left( \frac{\partial u_i u_j}
{\partial x_j} \right) \ 
&=\ -\frac{\partial p}{\partial x_i}\
+\ \frac{\partial}{\partial x_j}\frac{\partial u_i}{\partial x_j}\ ,\\ 
\frac{\partial u_i}{\partial x_i}\ &=\ 0\ ,
\end{split}
\end{align}
where $t$ is the dimensionless time, $x_j$'s are the dimensionless Cartesian 
coordinates ($j=1,2$ or $3$), $u_i$'s are the corresponding dimensionless velocity components, 
$p$ is the dimensionless pressure and $Re=\rho UL/\mu$ is the Reynolds number 
\cite{Devi02}.

\begin{figure}[h!]
\vspace{0.0in}
\centerline{
\epsfxsize=5.5in \epsffile{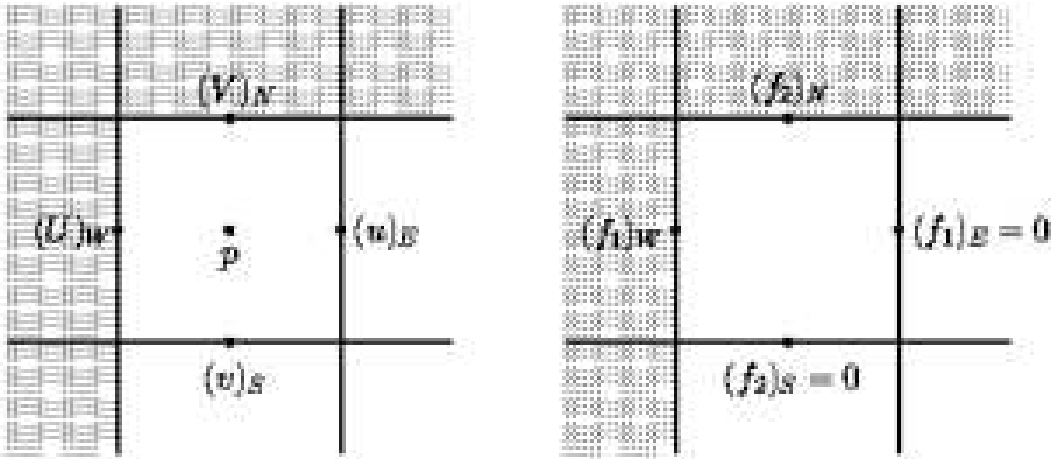}
}
\caption{Forcing points in staggered mesh.}
\label{fig:stagger}
\end{figure}

In the limit $Re \rightarrow 0$, we obtain the 
dimensionless unsteady Stokes flow equations, which is of central
concern in this study. These unsteady Stokes equations 
recover the steady Stokes formulation for infinite 
time advancement. Due to the numerical difficulties in generating complex 
meshes of various minimal surfaces, the immersed boundary finite volume method 
was adopted in this paper \cite{Jkim01,Fadl00}. The discrete-time momentum
forcing, $f_i$, is applied to satisfy the no-slip condition on the immersed
boundary. The forcing points are located in a staggered fashion like the 
velocity components defined on a staggered grid (see Fig.~\ref{fig:stagger}).
Therefore, the governing equations for unsteady incompressible stokes flow 
becomes
\begin{align}\label{eq:stokes}
\begin{split}
\frac{\partial u_i}{\partial t}\ &=\ -\frac{\partial p}{\partial x_i}\
+\ \frac{\partial}{\partial x_j}\frac{\partial u_i}{\partial x_j}\ + f_i ,\\ 
\frac{\partial u_i}{\partial x_i}\ &=\ 0\ ,
\end{split}
\end{align}
where $f_i$'s are the dimensionless momentum forcing components 
defined at the cell faces on the immersed boundary and inside the body. 

The Stokes equations represented by Eq. (\ref{eq:stokes}) are advanced with a fully implicit, fractional
step method \cite{Jkim01,Le91} with the Crank-Nicholson method in convection 
term. The fractional step, or time-split method, approximates the evolution
equation by decomposition of the operators in it.
A four-step time advancement \cite{Choi94} is modified for time integration 
of Eq. (\ref{eq:stokes}) as follows: 
\begin{align}
\label{eq:evolv}
\frac{\hat{u}_i - u^n_i}{\Delta t}\ &=\ -\frac{\partial p^n}{\partial x_i}\
+\ \frac{1}{2}\frac{\partial}{\partial x_j}\frac{\partial}{\partial x_j} 
(\hat{u}_i+u^n_i)\ + f^n_i\ ,\\ 
\frac{\bar{u}_i-\hat{u}_i}{\Delta t}\ &=\ \frac{\partial p^n}{\partial x_i}\ ,\\
\label{eq:poisson}
\frac{\partial}{\partial x_i}\frac{\partial p^{n+1}}{\partial x_i}\ &=\ 
\frac{1}{\Delta t} \frac{\partial \bar{u}_i}{\partial x_i}\ ,\\
\frac{u^{n+1}_i - \bar{u}_i}{\Delta t}\ &=\ -\frac{\partial p^{n+1}}
{\partial x_i}\ .
\end{align}
Here, the solution $u_i^n$ at time step $n$ is advanced  
to $u_i^{n+1}$ at time step $n+1$ through intermediate velocities $\hat{u}_i$ and $\bar{u}_i$.
Explicit direct forcing scheme for the momentum equation is utilized to obtain
the momentum forcing value in Eq. (\ref{eq:evolv}) which should be known 
{\it a priori} for time advancement \cite{Choi01}.
Discretizing Eq. (\ref{eq:evolv}) explicitly in time and rearranging it
for $f^n_i$ at a forcing point results in
\begin{equation}\label{eq:forcing}
f^n_i\ =\ \frac{\hat{U}_i - u^n_i}{\Delta t}\ +\frac{\partial p^n}
{\partial x_i}\ -\ \frac{\partial}{\partial x_j}\frac{\partial u^n_i}
{\partial x_j}\ , 
\end{equation}
where $\hat{U}_i$ is the velocity imposed at forcing point to enforce 
$u^n_i = \hat{U}_i$ on the immersed boundary. To satisfy the no-slip boundary
condition on the immersed boundary, $\hat{U}_i$ is zero when the forcing point 
is at the immersed boundary or inside the body.
All the spatial derivatives are resolved with the second-order 
central-difference scheme using a staggered mesh system. The discretized 
momentum equations are factorized into three operators which requires inversion
of tridiagonal matrices. Time advancement is achieved by sweeping $x_1$-, 
$x_2$-, and $x_3$-direction tridiagonal matrix algorithm at time intervals of
$n+\frac{1}{3}$, $n+\frac{2}{3}$, and $n+1$. 

To drive flow in a unit cell of a triply periodic porous medium, 
a pressure drop is applied from the inlet to the outlet of the channel in
one of the three directions ($x_1$ in this paper). In the other two 
directions ($x_2$ and $x_3$), standard treatment for velocity and pressure
periodic boundary conditions are applied. 
To solve the elliptic pressure Eq. (\ref{eq:poisson}), the Alternating Direction
Implicit (ADI) method is applied with a relaxation.
Non-uniqueness of a pressure solution due to applying a pressure drop condition
can be resolved with a point constraint which imposes an arbitrary pressure 
value at an arbitrary position. In fact, the pressure gradient only is 
necessary to solve for velocity field.
When the triply periodic surface possesses simple cubic symmetry, the 
periodic boundary conditions can be exploited further. Because of the mirror
image of the flow and pressure field across $x_2 = 0.5$ and $x_3 = 0.5$ in
the unit size channel simulation, the simple cubic symmetry motivates setting 
the velocity and pressure at the reflection position across the middle 
cross-section.

In general, the forcing point of an arbitrary immersed body does not coincide
with the immersed boundary, but rather inside the body in the non-body fitted 
Cartesian mesh. This requires a velocity interpolation to impose the correct 
$\hat{U}_i$ at a forcing point.  However, no velocity interpolation scheme 
is necessary to satisfy the no-slip boundary condition on the immersed boundary
in this paper because in the digitized triply periodic surfaces the forcing 
point of an immersed body always coincides with the immersed boundary.  

\section{Numerical Results}
\label{section4}

In this section, we will apply the immersed boundary method to determine Stokes flow
velocity fields in triply periodic porous media  with cubic symmetry.
We begin by computing the flow field for a periodic array of spheres in 
order to validate the procedure and the numerical results. Subsequently, we apply the
numerical procedure to two different pore-channel models
as well as the bicontinuous structures with interfaces that are the
Schwartz P and D minimal surfaces and the Schoen G minimal surfaces.
The permeability results for all six cases are compared at the
porosity $\phi = 1/2$. The fluid permeabilities are obtained from the imposed 
pressure drop $\Delta p$ and average velocity $\bar{U}$ of the flow field in 
the cubic periodic cell of side length $L$. Because of the cubic symmetry,
the second-order fluid permeability tensor is isotropic (i.e., it
is given by $k {\bf I}$, where $\bf I$ is the identity tensor)\cite{To02a}, and therefore
the determination of the permeability along the $x_1$ axes is sufficient
to characterize $k$. From Eq. (\ref{eq:darcy}), the fluid
permeability $k$ can be made dimensionless by dividing
it by $L^2$. Henceforth, all permeabilities will be reported in this
dimensionless form.

The computational domain is chosen to be a unit cube and staggered Cartesian 
meshes were generated, depending on the required resolution. All
initial velocities and pressures are set equal to zero. 

\subsection{Periodic array of spheres}

To verify the numerical accuracy of the immersed boundary method, we compare
our calculations to the analytical/numerical results of Sangani and Acrivos 
\cite{Sang82} for flow past periodic arrays of spheres centered on the sites of a simple
cubic lattice. The immersed body (sphere in this case) is constructed using
the following criterion. The immersed sphere is identified if the center of 
each voxel constituting a single sphere centered at ($0.5,0.5,0.5$)
satisfies the equation
\begin{equation}\label{eq:sphere}
(x_1-0.5)^2 + (x_2-0.5)^2 + (x_3-0.5)^2\ \le\ r^2\ , 
\end{equation}
where $r$ is the radius of the sphere.
As explained in Section \ref{section3}, a pressure drop is applied in $x_1$ direction, 
and standard treatment for velocity periodic boundary conditions are applied in
$x_2$ and $x_3$ direction. The fluid permeability $k$ obtained 
from numerical simulations  are listed in Table 
\ref{table:1} for various values of the scaled particle volume fraction $\chi= (c/c_{max})^{1/3}$,
where $c$ is the particle volume fraction and $c_{max}=\pi/6$ is the maximum
particle volume fraction corresponding to the case when the spheres 
are in contact. Our  fluid permeability calculations are  in good agreement 
with the results of Sangani and Acrivos \cite{Sang82}.

\begin{figure}[h!]
\vspace{0.5in}
\centerline{
\epsfxsize=5.0in \epsffile{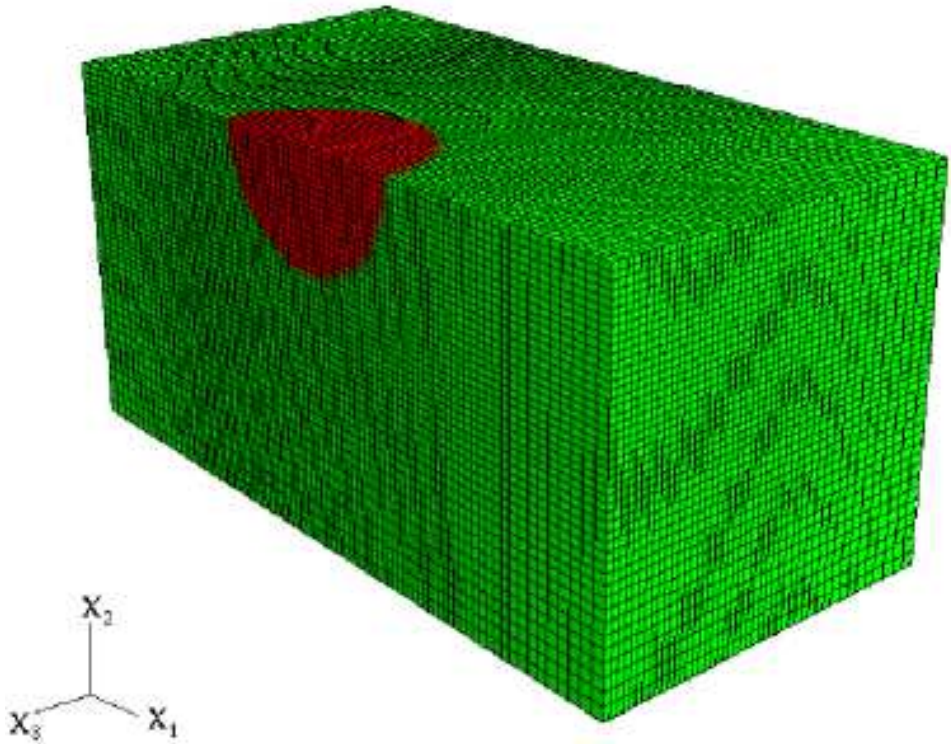} 
}
\caption{(Color online) Mesh for flow field over a sphere.}
\label{fig:mesh}
\end{figure}

\begin{figure}[h!]
\centerline{
\epsfig{file=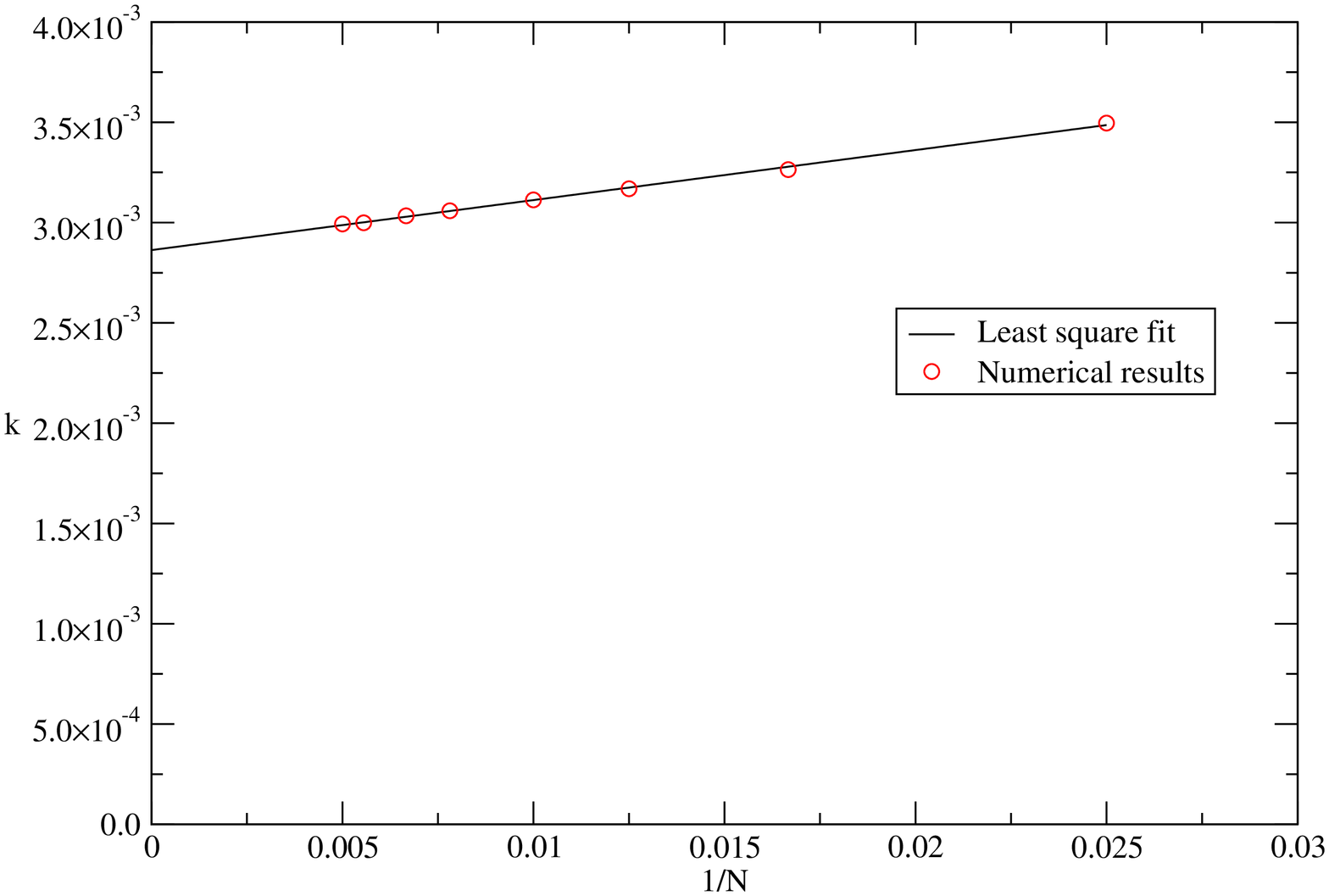,height=5.5in,angle=-90} 
}
\vspace{-0.1in}
\caption{(Color online) Fluid permeability $k$ for Stokes flow past a simple cubic array of 
equi-sized spherical obstacles as a function of $1/N$ for $\phi=1/2$.}
\label{fig:resolution}
\end{figure}

Non-uniform (stretching) staggered meshing is used
toward the body over the computational domain to get the required spatial 
accuracy for the immersed boundary method, which is 
important for comparison of our permeability data to corresponding analytical/numerical 
results over a smooth sphere (see Fig.~\ref{fig:mesh}). 
The discretization of a continuous body with a finite resolution obviously affects
the numerical accuracy of the computed fluid permeability $k$, which
depends on the value of the grid mesh spacing $h$  or the 
mesh resolution $N=1/h$. We studied the resolution effects on
the numerical accuracy at fixed porosity $\phi = 1/2$. The fluid permeability 
data and a corresponding linear least-squares fit are shown against $1/N$ in Fig. 
\ref{fig:resolution}.

Various mesh resolutions ($N = 20, 40, 60, 80, 128, 150, 180, 200$) are used to 
reduce the digitization error over solid regions in these simulations.
Generally, fluid permeabilities determined from the
immersed boundary method for digitized spheres are overestimated  but
this error, as Table \ref{table:1} shows,  decreases as the resolution increases.

\begin{table}[h]
\vspace{0.0in}
\noindent
\small
\caption{The fluid permeability $k$ for Stokes flow past simple cubic lattice
of spheres. Here $\chi= (c/c_{max})^{1/3}$,
where $c$ is the particle volume fraction and $c_{max}=\pi/6$.}
\begin{tabular}{|c|c|c|} \hline
\makebox[0.75in]{$\chi$} & \makebox[1.75in]{$k$ (Present work)} & 
\makebox[1.75in]{$k$ (Sangani \& Acrivos)} \\ 
\hline
 0.1  & $9.1467 \times 10^{-1}$ & $9.1107 \times 10^{-1}$ \\ 
 0.2  & $3.8436 \times 10^{-1}$ & $3.8219 \times 10^{-1}$ \\ 
 0.3  & $2.1069 \times 10^{-1}$ & $2.0805 \times 10^{-1}$ \\ 
 0.4  & $1.2516 \times 10^{-1}$ & $1.2327 \times 10^{-1}$ \\ 
 0.5  & $7.6379 \times 10^{-2}$ & $7.4668 \times 10^{-2}$ \\ 
 0.6  & $4.5803 \times 10^{-2}$ & $4.4501 \times 10^{-2}$ \\ 
 0.7  & $2.6059 \times 10^{-2}$ & $2.5246 \times 10^{-2}$ \\ 
 0.8  & $1.3758 \times 10^{-2}$ & $1.3197 \times 10^{-2}$ \\ 
 0.85 & $9.5964 \times 10^{-3}$ & $9.1516 \times 10^{-3}$ \\ 
 0.9  & $6.4803 \times 10^{-3}$ & $6.1531 \times 10^{-3}$ \\ 
 0.95 & $4.2473 \times 10^{-3}$ & $4.0031 \times 10^{-3}$ \\ 
 1.0  & $2.6727 \times 10^{-3}$ & $2.5203 \times 10^{-3}$ \\ 
\hline
\end{tabular}
\label{table:1}
\vspace{0.3in}
\end{table}

\subsection{Bicontinuous Triply Periodic Porous Media}

\begin{figure}[bthp]
\centerline{
\epsfxsize=2.5in \epsffile{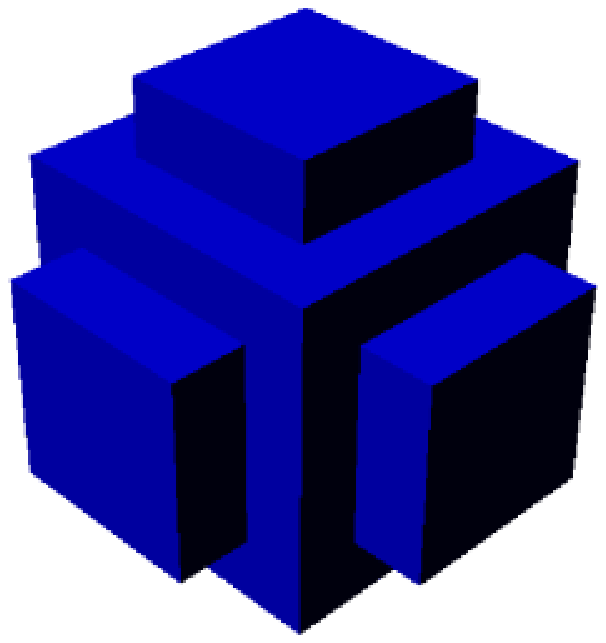} \hspace{0.1in}
\epsfxsize=2.5in \epsffile{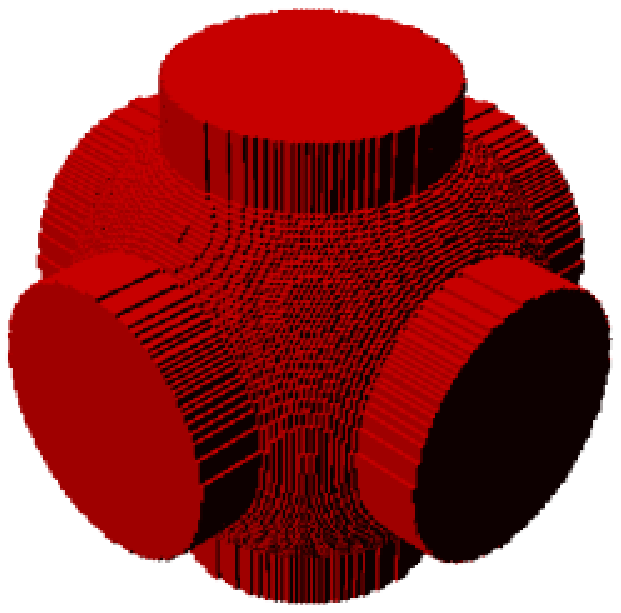}
}
\caption{(Color online) Unit cells of two different pore-channel models.
Left panel: cubic pore-square channel. Right panel:
spherical pore-circular channel.}
\label{fig:channel}
\end{figure}
\begin{figure}[h]
\vspace{0.0in}
\centerline{
\hspace{0.0in} \epsfxsize=5.0in \epsffile{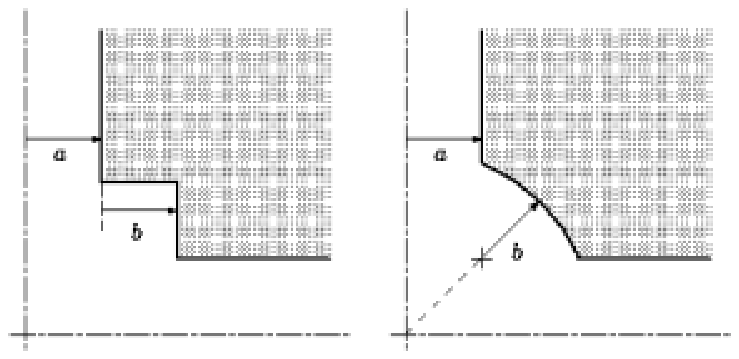}
}
\caption{Cross-sections of the two different pore-channel models. }
\label{fig:crosect}
\end{figure}

Before describing our results for the minimal surfaces, we first
consider two different pore-channel models that
are also bicontinuous and triply periodic. The latter is motivated
by the fact that one-dimensional flow inside channels are characterized by
high permeabilities. However, since we are interested in macroscopically
isotropic permeabilities (i.e., cases when the fluid permeability tensor
is isotropic), we consider triply periodic channels that intersect
in three orthogonal directions ($x_1$, $x_2$ and $x_3$). 
A pore in this model is the region where the channels meet
and its geometry depends on the geometry of the channel.
In one model, we consider square channels and cubic pores.
In the second model, we consider circular channels and spherical
pores (see Fig.~\ref{fig:channel}). Cross-sections of the two models are shown 
in Fig.~\ref{fig:crosect}, where $a$ and $b$ are the parameters
that enable us to control the relative channel/pore geometries.
For example, in the case of the cubic pore-square channel or spherical pore-circular channel, 
the pore is defined to be {\it minimal} when $b=0$.

\subsubsection{Pore-Channel Models}

\begin{figure}[ht]
\centerline{
\epsfxsize=4.0in \epsffile{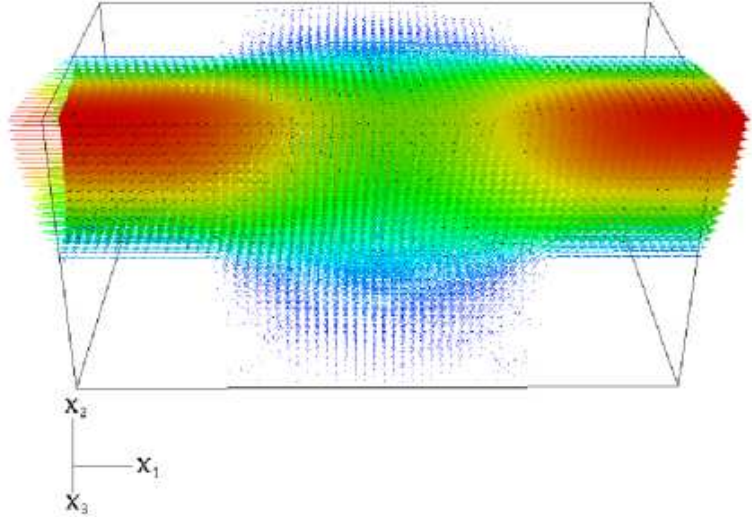}
}
\caption{(Color) The flow field in a unit cell of the triply periodic
 cubic pore-square channel
geometry for the minimal pore case ($a = 0.250$, $b = 0$) at
porosity $\phi=1/2$. Color scheme: red, yellow, 
green, light blue, and blue represent the spectrum of speeds from
highest to lowest, respectively.}
\label{fig:square_sect1}
\end{figure}

\begin{figure}[ht]
\centerline{
\epsfxsize=4.0in \epsffile{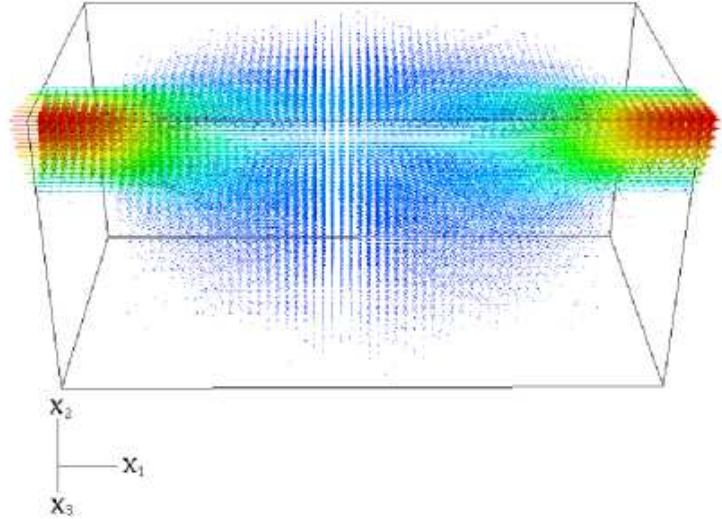}
}
\caption{(Color) The flow field in a unit cell of the triply periodic
 cubic pore-square channel
geometry for the case $a = 0.132$, $b = 0.250$ at
porosity $\phi=1/2$. The color scheme for speeds is the same
as described in Fig. \ref{fig:square_sect1}}.
\label{fig:square_sect2}
\end{figure}

We calculated the permeability $k$ for three  different combinations of $a$ and
$b$ at three porosities $\phi = 0.5, 0.7$ and $ 0.9$ on fine mesh of {\bf $128 \times 
128 \times 128$}. Table \ref{table:2} summarizes our permeability results for the
 cubic pore-square channel geometry. Not surprisingly, the permeability increases with 
increasing porosity $\phi$. The minimal pore case (i.e., $b=0$)
provided the highest permeability at any porosity. As shown in 
Fig.~\ref{fig:square_sect1}, channel flow away from the
channel intersection (both upstream and downstream) has a  Poiseuille-type flow profile 
but, in the vicinity of the channel intersection, the flow
is no longer confined and the associated maximum speed 
in the $x_2$-$x_3$ plane is reduced. This is illustrated in
Fig.~\ref{fig:square_sect1} where the  speed profile for
the minimal pore case.  Increasing the pore size
while keeping the porosity fixed, leads to an even lower maximum speed,
as depicted  in Fig.~\ref{fig:square_sect2}. Consequently, the pore region at the 
channel intersection has a significant effect on the overall flow 
field and hence fluid permeability of the porous media.

\begin{table}[h!]
\vspace{1.0in}
\noindent
\small
\caption{The fluid permeability for the  cubic pore-square channel model
for several porosites and geometric parameters $a$ and $b$.}
\begin{tabular}{|c|c|c|c|} \hline
\makebox[0.55in]{$\phi$} & \makebox[0.75in]{$a$} & \makebox[0.75in]{$b$} &
\makebox[1.35in]{$k$} \\
\hline
      &  0.2500  &  0.0000  & $3.0744 \times 10^{-3}$ \\
 0.5  &  0.1324  &  0.2500  & $0.5310 \times 10^{-3}$ \\
      &  0.0781  &  0.3125  & $0.0948 \times 10^{-3}$ \\
\hline
      &  0.3125  &  0.0000  & $9.1168 \times 10^{-3}$ \\
 0.7  &  0.2656  &  0.1406  & $6.2495 \times 10^{-3}$ \\
      &  0.2250  &  0.2000  & $3.7939 \times 10^{-3}$ \\
\hline
      &  0.4000  &  0.0000  & $26.207 \times 10^{-3}$ \\
 0.9  &  0.3750  &  0.0750  & $23.915 \times 10^{-3}$ \\
      &  0.2625  &  0.2125  & $8.0231 \times 10^{-3}$ \\
\hline
\end{tabular}
\label{table:2}
\end{table}

\begin{figure}[ht]
\centerline{
\epsfxsize=2.2in \epsffile{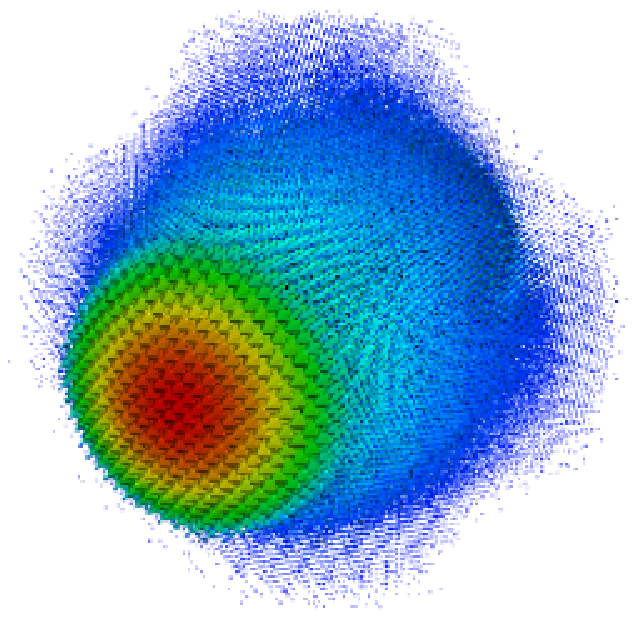} \hspace{0.0in}
\epsfxsize=2.1in \epsffile{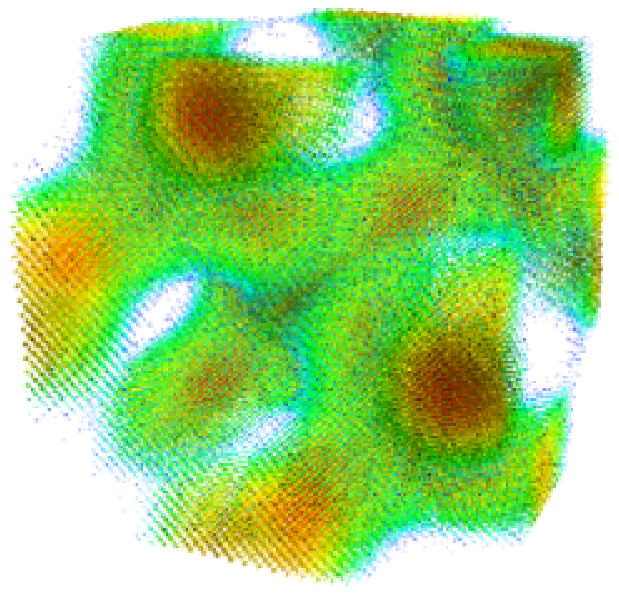} \hspace{0.0in}
\epsfxsize=2.1in \epsffile{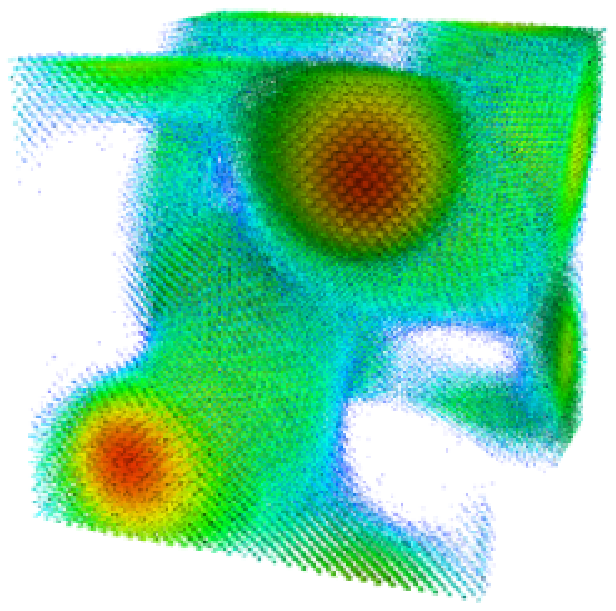} 
}
\caption{(Color) Flow fields for triply periodic porous media with 
interfaces that are minimal surfaces.
Left panel: Schwartz P surface.
Center panel: Schwartz D  surface.
Right panel: Schoen G surface.
The color scheme for speeds is the same
as described in Fig. \ref{fig:square_sect1}}.
\label{fig:tpmsflow}
\end{figure}

We also computed fluid permeabilities for the spherical pore-circular channel
geometry but since the trends are the same as in Table \ref{table:2},
we do not present all of these results here.
In the minimal pore case ($b=0$), the fluid permeability of the circular channel
geometry ($k=3.4596 \times 10^{-3}$) is larger than the corresponding 
square channel case  ($k=3.0743 \times 10^{-3}$).
Moreover, the circular channel has a lower specific surface than
the square channel at the same porosity.
This point will be discussed further below.

\subsubsection{Triply Periodic Minimal Surfaces}

Here we report results for the fluid permeability of triply
periodic porous media with interfaces that are the
Schwartz P and D surfaces and the Schoen G surface.
The minimal surfaces are generated from the  aforementioned 
Landau-Ginzburg potential minimization
and each has a fine mesh resolution of  $129 \times 129 \times 129$. 
The potential data $\psi(x)$ are numerically digitized such that $\psi(x)>0$ 
for points in phase 1 and $\psi(x)<0$ for points in phase 2 or vice versa.

It is found that the computational difficulty due to the geometrical
complexities are very well treated by the immersed boundary finite volume 
method. Representative flow fields over the  minimal surfaces are 
shown in Fig.~\ref{fig:tpmsflow}. 
Our computed fluid permeabilities for the three minimal
surfaces are summarized in Table \ref{table:3} along with corresponding
results for the circular and square channels with minimal pores
and the simple cubic array of spherical obstacles at the porosity $\phi=1/2$.
It should be noted that the Schwartz P surface has the highest fluid permeability
among these six triply periodic porous media. Included in the table is the 
specific surface $s$ for each of the structures. The fluid permeabilities  
are seen to be inversely proportional to the specific surfaces. Thus, the 
Schwartz P surface has the largest fluid permeability
followed by the circular channel, square channel, spherical obstacle, and finally
the Schoen G and the Schwartz D surface cases. 
The permeability of the circular channel  with a minimal pore 
is actually close in value to that of the Schwartz P surface.
Evidently, the sharp corner present in the former results
in slightly greater energy dissipation than in the smooth Schwartz P surface.

\begin{table}[h!]
\vspace{0.0in}
\noindent
\small
\caption{The fluid permeability $k$ and specific surface $s$ at $\phi=1/2$ for six
different triply periodic porous media models.
The pore-channel results are presented for the minimal pore case ($b=0$).
SC and BCC stand for simple cubic and body centered cubic lattices,
respectively.}
\begin{tabular}{|c|c|c|c|} \hline
\makebox[0.95in]{Model} & \makebox[1.35in]{$k$} & 
\makebox[0.95in]{$s$} & \makebox[0.95in]{Symmetry} \\ 
\hline
 Schwartz P surface&  $3.4765 \times 10^{-3}$  & 2.3705 & SC  \\ 
 Circular channel/spherical pore  &  $3.4596 \times 10^{-3}$  & 2.6399 & SC  \\ 
 Square channel/cubic pore  &  $3.0743 \times 10^{-3}$  & 3.0000 & SC  \\ 
 Spherical obstacle     &  $3.0591 \times 10^{-3}$  & 3.0780 & SC  \\ 
 Schoen G  surface &  $2.2889 \times 10^{-3}$  & 3.1284 & BCC \\ 
 Schwartz D surface&  $1.4397 \times 10^{-3}$  & 3.9011 & SC  \\ 
\hline
\end{tabular}
\label{table:3}
\end{table}

\section{Conclusions and Discussion}
\label{section5}

We have determined the Stokes flow fields and associated fluid
permeabilities through six different triply porous
media:  the Schwartz P, the Schwartz D, and the Schoen G
minimal surfaces, two different pore-channel models, and a
simple cubic array of spherical obstacles.
The latter model served as a benchmark to verify the accuracy of our numerical
procedure: the immersed boundary finite volume method. We found good
agreement with previous results for the simple cubic array.
Interestingly, we found that the Schwartz P porous medium has the highest 
fluid permeability among all of the six triply periodic porous
media considered in this paper. The fluid permeabilities  
were shown to be inversely proportional to the specific surfaces for these bicontinuous structures. 
This might lead one to conjecture that the maximal fluid permeability (scaled
by the cell length squared) for
a triply periodic porous medium at a porosity $\phi=1/2$ is achieved
by the structure that globally minimizes the specific surface.
As we pointed out in the Introduction, the specific surface does not necessarily
correlate with the permeability. However, the pore spaces of
structures with large permeabilities are expected to be {\it simply
connected} and therefore it would not be unreasonable for the permeability
to be inversely  proportional to the specific surface in these instances.

The determination of the structure that globally optimizes the fluid
permeability is a highly challenging problem. It is only very recently
that optimal bounds on $k$ have been identified. By contrast, optimal bounds on the effective
conductivity and elastic moduli of two-phase composite media 
have long been known \cite{To02a,Mi02}.
In particular, Torquato and Pham \cite{To04b} showed that the so-called void
upper bounds on $k$ for flow in anisotropic media with a single preferred direction
are optimal for certain coated-cylinders models. Such structures arise
in practice  in sea ice, for example,
which has a microstructure that is well modeled by oriented cylindrical
channels \cite{golden}. However, these bounds are not applicable for our
present purposes because we would like to ascertain
the triply periodic structure that maximizes the isotropic permeability.
Even if the conjecture that the maximal dimensionless fluid permeability for
a triply periodic porous medium with a simply connected pore space
at a porosity $\phi=1/2$ is achieved
by the structure that globally minimizes the specific surface is true,
finding the minimizing specific surface structure is highly nontrivial.
For example, Kelvin's problem, the determination
of the space-filling arrangement of congruent {\it closed} cells of equal volume that
minimizes the surface area, is still an open question, although it is believed that
the Weaire-Phelan structure \cite{We94} is a good candidate.  
In a future work, we will determine whether the Schwartz P triply periodic
porous medium is a locally maximal structure for the isotropic fluid
permeability.

\begin{acknowledgments}

The authors are grateful to Volodymyr Babin and Wojciech G{\'o\'z}d{\'z} for 
providing TPMS data and to Jung-Il Choi for helpful discussions. This work was supported by the Air Force Office of Scientific Research
under Grant No. F49620-03-1-0406.
\end{acknowledgments}

\end{document}